# Towards Measuring the Adaptability of an AO4BPEL Process

Khavee Agustus Botangen
School of Engineering, Computer and
Mathematical Sciences, Auckland
University of Technology,
New Zealand
kbotange@aut.ac.nz

Jian Yu
School of Engineering, Computer
and Mathematical Sciences,
Auckland University of Technology,
New Zealand
jian.yu@aut.ac.nz

Michael Sheng
Macquarie University,
Australia
michael.sheng@mq.edu.au

## ABSTRACT

Adaptability is a significant property which enables software systems to continuously provide the required functionality and achieve optimal performance. The recognised importance of adaptability makes its evaluation an essential task. However, the various adaptability dimensions and implementation mechanisms make adaptive strategies difficult to evaluate. In service oriented computing, several frameworks that extend the WS-BPEL, the *de facto* standard in composing distributed business applications, focus on enabling the adaptability of processes. We aim to evaluate the adaptability of processes specified from the extended-BPEL frameworks. In this paper, we propose metrics to measure the adaptability of an AO4BPEL process. The metrics is grounded in the perspective that a process is capable of dynamically adapting to changes in business requirements. This opens potential future work on evaluating the adaptability of processes specified from various aspect-oriented WS-BPEL frameworks.

## 1. INTRODUCTION

Adaptability has been one of the challenges in the design of software systems [10, 15]. The ever-changing user requirements, the volatile environment, and the pervasive accessibility of systems ascertain that adaptability is a desirable quality. It allows a system to adjust itself to changes and continue to offer the required functionality and quality of service. Hence, several works are focused on engineering adaptable computing paradigms to incorporate techniques for handling the necessary flexibility of systems [12].

Service oriented computing is a paradigm that integrates web services to create distributed systems. The integration, which results to a *composite service*, allows organisations to build a new application from multiple heterogeneous functionalities that are exposed as services. Among various technologies, Web Services Business Process Execution Language (WS-BPEL or BPEL) [3] has become a *de facto* standard to specify composite services [18]. The language describes the aggregation of services *a.k.a.* *orchestration* or *service composition* that satisfies an underlying business process flow. BPEL is strongly supported by industry and the open-source community; a number of BPEL engines have been developed such as Oracle's BPEL Process Manager, Microsoft's BizTalk Server, IBM's WebSphere Process Server, SAP's Exchange Infrastructure, and the Apache ODE. However, BPEL does not provide appropriate support for the dynamic adaptation of processes [2]. Hence, the need for process adaptability has attracted several research works to extend BPEL's basic functionality. In [18], the authors enumerated several extended BPEL frameworks that aimed for process adaptability. These frameworks implement adaptability mechanisms such as: aspect injection, message interception, late binding, or explicit integration –this is to enhance a BPEL process' adaptation capability.

With adaptation, a composite service can be efficiently extended, changed, customised, or configured for a particular context of use [13]. Furthermore, adaptability contributes to achieving vital quality attributes such as: (re)usability [5], portability [8], maintainability [6], scalability and evolution of processes. Recognising the significance of adaptability to overall system quality, the ultimate goal is its design, implementation, and evaluation in the life cycle of service compositions. The initial step we put forward to achieve this goal is to be able to quantify adaptability. A concrete adaptability measure would enable a designer to compare processes and to decide on quality compositions. However, adaptability solutions have a wide range of objectives and implementation mechanisms. This leads to various adaptability dimensions and interpretations which make adaptive strategies difficult to evaluate.

In this paper, we propose metrics to measure the adaptability of a process[1] that is specified from an AO4BPEL [2] framework. The popular AO4BPEL framework extends the standard BPEL through aspect injection to realise runtime process adaptability. With our aim to evaluate the adaptability of aspect-oriented BPEL processes, we examine the implementation of AO4BPEL to formulate our approach. There are approaches introduced to measure software and process adaptability, such as [14] and [11], which can compute adaptability values based on the predefined composition logic and the number of comprising components. These approaches quantify adaptability of static process structures. However, their assumption excludes the perspective of systems in a highly dynamic context where it is necessary to have structure-variable processes. Indeed, our proposed metrics consider a flexible process structure that is implemented in the AO4BPEL framework.

The rest of the paper is structured as follows. In Section 2, we briefly discuss the related work. In Section 3, we introduce the aspect-oriented BPEL mechanism and the concept of adaptability that underpins our approach. Section 4 presents the proposed metrics. We demonstrate the use of the metrics in Section 5 and present our conclusion and future work in Section 6.

## 2. RELATED WORK

Despite several adaptability metrics found in the literature, we only discuss the most related work to our approach. The authors in [14] present a metrics to measure adaptability considering the number of components that provide or can provide the services/functionalities comprising a software architecture. The weighted adaptability index of each service is aggregated to get the overall adaptability of the software. Moreover, they propose that adaptability should be monitored against other system quality attributes (e.g. availability, cost, and response time) because of possible trade-offs. Their perspective in measuring adaptability is to be able to choose the most adaptable architecture that fulfils the required level of a certain quality attribute.

On measuring adaptability of a process, the authors in [11] extend the metrics in [14] for BPEL processes. They compute the overall adaptability of a process by aggregating the derived adaptability indexes of its elements (i.e. activities). The element adaptability index is based on the number of concrete partner services and the element's behaviour. Indeed, they can measure the adaptability of a process to select its concrete partner services. This assumption is different from our work because they consider a static BPEL structure which becomes variable through bindings to potential partner services specified at design time.

Lenhard [8] proposed a metrics which is further expanded in [9] to quantify the degree of structural adaptability of processes to satisfy code portability. The purpose is to measure the likelihood of a process to be adapted in a different form considering portability to another execution platform. Hence, the primary dimension for adaptability measurement is the amount of alternative representations that a language offers for a particular element; the more alternatives for the process elements, the easier to modify the process to be ported to a new runtime platform. This also relates to code-level adaptability because the more alternatives that exist for a process element, the easier it could be replaced with the alternatives supported by a runtime platform, and hence, the more adaptable the resulting code.

In [6], adaptability of a process is perceived as the flexibility to adjust to new, different, or changing requirements. The authors argue that context-independency (i.e. degree of coupling) of a process to its environment (e.g. partner web services, clients, and resources) makes a composite service adaptable. Thus, they propose a metrics to measure the context-independency of a BPEL process. In [5], adaptability is implied through the concept of reusability. The authors assume the volatility of business rules; whenever business rules change, a process may change accordingly. A composite service should be able to adapt to the future changes in requirements and business rules to become (re)usable. Hence, the proposed metrics either quantify or estimate the potential reusability of a BPEL process.

In contrast, our work considers adaptability as the degree to which a BPEL process structure becomes variable with regards to expected runtime changes in workflows to expose new process behaviour. This adaptability perspective has not been considered in the existing metrics and is a common focus of the extended BPEL frameworks.

## 3. ADAPTABILITY IN AN ASPECT-ORIENTED BPEL PROCESS

Implementing adaptation mechanisms in service compositions is classified in several ways [18]. Some general classifications include the following. *Manual* adaptation needs human effort (e.g. inserting codes) to instruct a system to adapt, while *automatic* adaptation is performed by the system based on predefined adaptation conditions (e.g. replacing a non-available service with its alternative). There are also works focused on adaptability to ensure *non-functional* quality requirements of a process, for example, to automatically replace a low-performing partner service with a better one. In contrast, *functional* adaptability copes with the changes in business requirements and environment context. *Proactive* and *reactive* adaptability approaches depend on whether adaptation happens before or after an event. The various approaches that enhanced the BPEL, commonly aim for *dynamic* adaptability of processes. Dynamic adaptability (*a.k.a. runtime adaptability*) happens when the behaviour of a process or some part of it changes as it executes without stopping or restarting it; as opposed to *static* adaptation where humans are needed to (re)configure the process and then restart it [1]. According to [2], the only way to implement runtime changes in BPEL is to stop a running process, modify the composition, and restart. However, there are critical systems that cannot be abruptly shut down to implement changes. Thus, works on adopting the aspect-oriented programming (AOP) [7] paradigm for extending the BPEL language and/or engine have been proposed (e.g. [2], [4], and [18]).

Integrating the aspect-oriented mechanism to BPEL aims to efficiently handle the runtime adaptation that is required by compositions to cope with changes. This increases flexibility, fosters reuse, and introduces a high degree of modularity and configurability in processes [4]. We examine AO4BPEL [2], this being one of the popular aspect-oriented BPEL implementations, to conceptualise our metrics. AO4BPEL classifies changes into *non-functional* (e.g. auditing, authentication, and logging) and *functional* crosscutting concerns (e.g. change in business rules that will change the business process flow logic). A crosscutting concern is encapsulated into an *aspect* which is dynamically weaved into a process. This introduces structure variability which allows a process' behaviour to be adapted at runtime. In this section, we introduce BPEL and its constructs; and we define the adaptability dimension that has become the basis of our metrics.

---

[1] We use the terms: *process*, *BPEL process*, *service composition*, *composite service* interchangeably throughout the paper.

## 3.1 The BPEL Process

BPEL is a workflow-based web service composition language that specifies, through its constructs/activities, the control flow and the data flow among partners (i.e. clients and web services) interacting with the composition [2], [4]. The control flow and data flow describe respectively the ordering of interactions and the data exchange among the partners. The resulting specification, which is also called BPEL process, is deployed and executed on a BPEL-compliant orchestration engine.

The BPEL process is composed of activities categorised as basic and structured. The basic activities are atomic, include <invoke>, <receive>, <reply>, and <assign> which are used respectively to invoke a partner web service, receive a message from client, generate response in a synchronous operation, and manipulate data variables. Structured activities determine the execution flow among the basic and structured activities that they contain. The structured activities <sequence>, <switch>, and <while> define the basic sequence control of activities; the <flow> activity defines synchronisation and concurrency; and the <pick> activity determines a choice based on an external event (i.e. message or alarm event).

```
1   <process name="TravelBooking" … >
2    <partnerLinks>
3      <partnerLink name="airline" … />
4        …
5    </partnerLinks>
6    <variables>
7      <variable name="clientrequest" … />
8        …
9    </variables>
10   <sequence name="mainSequence">
11     <receive name="receiveClientRequest"… />
12     <assign>… </assign>
13     <invoke name="invokeAirlinesService"…
14          … operation="bookFlight" />
15     <invoke name="invokeHotelsService" …
16          … operation="bookHotel" />
17     <assign>… </assign>
18     <reply name="responseToClient" … />
19   </sequence>
20  </process>
```

**Figure 1. A travel booking process.**

As an example, consider a simplified travel booking process that arranges both the flight and accommodation booking for a client. The process connects to the Airlines Web Service which provides the flight booking function and the Hotels Web Service which performs accommodation booking. It takes as input the flight details and client details, and returns a string description of the travel package. Figure 1 shows a BPEL process outline of the booking process. *Partner links* (lines 2-5) define the different parties that interact with the process. The three partner links could be defined respectively: one for the client that invokes the process and the other two for the web services invoked by the process. The defined *variables* (lines 6-9) are used to store, reformat, and transform messages sent to and received from partners. Each variable has a message type that is defined in the WSDL file of the composition or one of its partners. The process main body (lines 10-19) is a sequence of activities: *<receive>*, which waits for a travel request message (i.e. contain the process inputs) from the client; *<assign>*, which copies needed data into variables as input to the web services defined in the *<invoke>* activities; and another *<assign>* activity which transforms the outputs of the invoked web services into a message which is sent to the client through the *<reply>* activity.

## 3.2 Process Variability

The adaptability we consider here is the ability of a process to be modified at runtime that leads to a variable-structure BPEL process. Modifications become necessary to address business workflow variations and crosscutting concerns. For instance, AO4BPEL introduces the capability of adding, excluding, or replacing process activities. This capability is defined in the aspects that encapsulate the crosscutting concerns. The places in the process where crosscutting concerns can be specified are generally called *join points*. An aspect describes the *pointcut* –a specific join point, the *advice* –new activity to realise a crosscutting concern, and the *advice type* –the timing of execution of the advice with regards to the occurrence of the pointcut. The advice is specified to a pointcut as *Before*, *Around*, or *After* activity. *Before* and *After* advice types execute respectively before and after the pointcut. An *Around* advice is interpreted as a replacement activity for the pointcut. We illustrate the AO4BPEL concepts in the simplified process representation shown in Figure 2. The process is composed of five activities *a*, *b*, *c*, *d*, and *e*. Each activity can be a join point for crosscutting concerns that can be defined in the aspects. Activity *a* is specified as a pointcut for the aspect $Q_1$; when the process executes, activities defined in the advice of $Q_1$ are executed *Before* performing *a*. Likewise, aspect $Q_2$ has an *Around* advice to be executed instead of the specified pointcuts *b* and *c*.

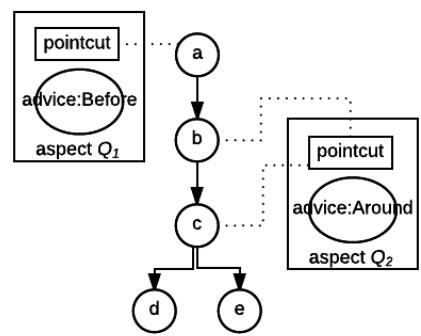

**Figure 2. A process with specified aspects.**

We take for example the travel booking process. Suppose a new requirement arises: to check the validity of a client's travel request (e.g. verify if the request has valid departure and arrival dates) before invoking the Airlines Web Service. AO4BPEL implements this change as an aspect. Figure 3 shows an outline of the *VerifyRequest* aspect. This aspect captures the occurrence of the invoke operation *bookFlight* as the pointcut (lines 8-11). The advice, which invokes operation *verify* of the Verifier Web Service, is a *sequence* activity that executes before the join point activity (lines 12-19). Hence, changes are encapsulated as aspects that are separate from the process. If a business rule changes, appropriate aspects can be activated or deactivated dynamically while the respective process is running, thus applying the adaptation at runtime.

```
1   <aspect name="VerifyRequest" … >
2     <partnerLinks
3       <partnerLink name="verifier" … />
4     </partnerLinks>
5     <variables>
6       ….
7     </variables>
8     <pointcut name="crosscut1"… >
9       //process[@name="TravelBooking"]
10      //invoke[@operation="bookFlight"]
11    </pointcut>
12    <advice type="before">
13      <sequence>
14        <assign>… </assign>
15        <invoke partnerLink="verifier"…
16              … operation="verify" />
17        <assign>… </assign>
18      </sequence>
19    </advice>
20  </aspect>
```

**Figure 3. The VerifyRequest aspect.**

In this regard, if an activity *a* in a process is a potential join point, we can specify to *a* the three types of advice (i.e. *Before*, *Around*, *After*). We interpret this as: *a* has three *variabilities*. This makes *a* a variable activity. The more variabilities specified to *a*, the easier *a* can be modified and adapt to possible runtime changes. Overall, the more variabilities a process has, the higher its adaptability.

## 4. MEASURING ADAPTABILITY

To measure process adaptability, we tag each join point activity with an adaptability value [16], [14], [11], [8]. This value depends on the number of specified variability (i.e. advice) to an activity. Then the atomic values are combined to get the adaptability degrees of one or more higher levels of structured activities, until the entire process is considered as a whole.

**Definition 1:**

We define the *Variability Value* $VV<a>$: is the cardinality of the set of variabilities $\{vr<a>_1… vr<a>_n\}$ specified for a join point activity *a*.

$VV<a> = |\ \{vr<a>_1… vr<a>_n\}\ |$, where *a* is a variable activity in the process and $vr<a>$ is a variability for *a*.

The absolute values should be mapped into a common range interval that will provide a standard and clear interpretation of values. We map the variability values into a percentage scale within the interval [0,...,1]. A variability value mapped to 0 means no adaptability while a value mapped to 1 means full-adaptability. For the values in between, a mapping value approaching 1 indicates better adaptability, while a mapping towards 0 means otherwise. We use a *reference value* ($R$) [9], which refers to the maximum variability value for an activity. $R$ will be determined by designers so that the most adaptable process activity is mapped to a value 1. The author in [8] mentions the importance of using $R$ in transforming absolute values into the interval counterpart values. When aggregating individual adaptability values, the resulting metrics will always be in the range [0,...,1]. Hence, by carefully setting a value for $R$, adaptability of different processes can be compared through the generated metrics.

**Definition 2:**

We define *Variability Degree* $VD<a>$: is the $VV$ of *a* relative to the value of $R$. If we assume the maximum number of advice for a join point activity is three (i.e. *Before*, *Around*, and *After*), in this case, $R$=3.

$VD<a> = VV<a>\ /\ R$, where $VV<a>$ is the *Variability Value* of activity *a* and $R$ is the maximum number of advice that can be specified for *a*.

The individual $VD$ of the variable activities are aggregated to get the concrete measure for the whole process. The aggregation should consider the behaviour of each BPEL structured activity construct that integrates the activities [11], [5], [6]. Thus, we define variability degree ($VD$) computations for the following structured activities and consider each formulation in the aggregation towards finding the $VD$ of a process.

**Switch activity** – a <*switch*> construct is composed of branches which are specified through *case* elements. Each branch defines a conditional behaviour that is chosen to execute depending on the Boolean value defined by its case element. A branch executes an activity within it when the condition of its case is true. When none of the cases is true, the optional branch which has the *otherwise* element is executed.

$VD<switch> = \sum_{i=1}^{n} (VD_{<a_i>} \cdot \frac{1}{n})$, where *n* is the number of activities and $a_i$ the *i*th activity contained in the <*switch*> construct.

The variability degree of the <*switch*> construct depends on the probability of each conditional branch to be executed. Only one conditional branch is executed per <switch> instance, hence we assume a similar probability for each branch: $\frac{1}{n}$ (i.e. *n* is the number of branches).

**Sequence activity** – a <*sequence*> construct contains activities that are to be executed in sequential order.

$VD<sequence> = (\sum_{i=1}^{n} VD_{<a_i>})\ /\ n$, where *n* is the number of variable activities and $a_i$ is the *i*th variable activity contained in the <*sequence*> construct.

The variability degree of the *<sequence>* construct is the arithmetic mean of the *VD* of all variable activities within it.

**Pick activity** – within a *<pick>* construct are branches which have activities associated to events (i.e. timer event or message event). A branch is executed on the occurrence of an event. Hence, the *<pick>* construct has a conditional behaviour similar to the *<switch>* construct but is based on events. In simultaneous events, only the first event to occur is processed. Generally, the pick activity defines two branches. The OnMessage branch is similar to the *<receive>* activity; it is used when a message is needed coming from an external entity. The OnAlarm branch is used when specifying an expiry time.

$$VD<pick> = \sum_{i=1}^{n} \left( VD_{<a_i>} \cdot \frac{1}{n} \right),$$ where $n$ is the number of activities and $a_i$ is the $i$th activity contained in the *<pick>* construct.

Similar to the *<switch>* construct, the variability degree of the *<pick>* construct depends on the probability of events to occur. Only one event should be processed per *<pick>* instance, hence we assume the probability: $\frac{1}{n}$ (i.e. $n$ is the number of events within the *<pick>* construct).

**Flow activity** – a *<flow>* construct executes its activities in parallel. It allows a BPEL process to perform multiple tasks at the same time. It also offers the possibility of synchronisation among the enclosed activities.

$$VD<flow> = \left( \sum_{i=1}^{n} VD_{<a_i>} \right) / n,$$ where $n$ is the number of variable activities and $a_i$ is the $i$th variable activity contained in the *<flow>* construct.

Similar to the *<sequence>* construct, all activities within the scope of the *<flow>* construct are executed. However, concurrent execution of activities are done in the latter.

**While activity** – a *<while>* construct defines the iterative execution of the enclosed activities. The activities within *<while>* are repeatedly executed while the specified Boolean condition holds true.

$$VD<while> = \left( \sum_{i=1}^{n} VD_{<a_i>} \right) / n,$$ where $n$ is the number of variable activities and $a_i$ is the $i$th variable activity contained in the *<while>* construct.

We treat the *<while>* construct similarly to the *<flow>* and *<sequence>* constructs. Despite the iterative execution of activities within its scope, the variability of the activities per execution will be the same.

**Definition 3:**

The aggregation for the root structured activity – usually a *<sequence>* – that contains the BPEL process' main body derives the *Process Adaptability Metric* (*PAM*).

$PAM_P = VD<sequence_{root}>$ where *<sequence_{root}>* is the root structured activity of process $P$.

In Figure 4, we show a simplified diagram of a BPEL process with six activities. The atomic activities *b*, *d*, *e*, and *f* have individual variability degrees. However, the structured activities *c* and *a* derive their variability degrees from the aggregation of the *VD* of elements they contain. For example, we aggregate $VD<e>$ and $VD<f>$ to derive $VD<c>$. Then, $VD<a>$ is the aggregation of $VD<b>$, $VD<c>$, and $VD<d>$. Activity *a* happens to be the root node, hence, $PAM_P = VD<a>$.

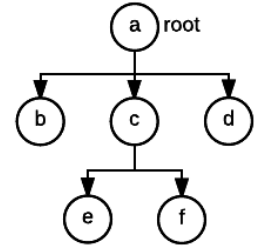

**Figure 4. A simplified BPEL process.**

## 5. APPLYING THE METRICS

We redefine the travel booking process we described in Section 3.1 to demonstrate the proposed metrics. We show in Figure 5 process $P$: the business logic, which is expressed in Business Process Modeling Notation (BPMN) [17], and its equivalent BPEL process, which we visualise through an activity tree graph. The process starts when a customer makes a request for travel booking. The request contains the travel and customer details. Based on the specified destination from the request, the process will either book a domestic or international flight. Likewise, the process books accommodation, then, returns a proposed travel package to the customer. The process is composed of $n$ tasks $t_i | i = 1, …, n$ and each task is associated with a partner web service $s_i | i = 1, …, n$ to be orchestrated by the process to perform the required functions. $P$ has five variable activities regarded as join points. We represent the variable activities with dashed borderlines. We take this assumption from the implementation of AO4BPEL in [2] where an advice can be specified to BPEL's messaging activities $<invoke>$, $<receive>$, and $<reply>$. The variabilities $\{v1, …, v6\}$ are specified to some variable activities. We assume these variabilities are defined in aspects containing the advice to handle changes in business rules.

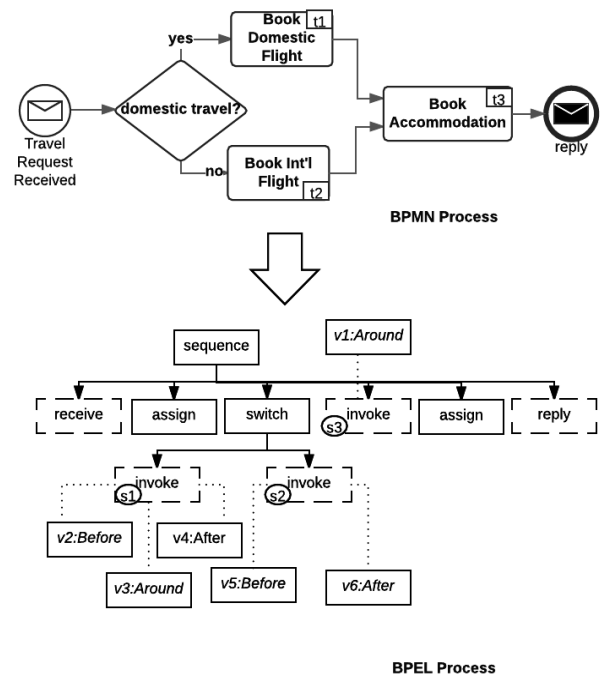

**Figure 5. Illustration of the travel booking process.**

We apply the metrics to derive the adaptability of $P$. First, we get the variability values $VV$ of the variable atomic activities in $P$:

$[VV<receive>, VV<invoke_{s1}>, VV<invoke_{s2}>, VV<invoke_{s3}>, VV<reply>] = [0, 3, 2, 1, 0]$.

Second, we map each individual $VV$ to a variability degree $VD$ by using a reference value $R$. In this case, $R=3$, maintaining our assumption that the maximum variabilities for a join point is three. We compute:

$[VD<receive>, VD<invoke_{s1}>, VD<invoke_{s2}>, VD<invoke_{s3}>, VD<reply>] = [VV<receive> / 3, VV<invoke_{s1}> / 3, VV<invoke_{s2}> / 3, VV<invoke_{s3}> / 3, VV<reply> / 3] = [0/3, 3/3, 2/3, 1/3, 0/3] = [0, 1, 0.67, 0.33, 0]$.

Third, we aggregate the individual variability degrees using the structured constructs:

$VD<switch>$

$= (VD<invoke_{s1}> * ½) + (VD<invoke_{s2}> * ½)$

$= (1 * 0.5) + (0.67 * 0.5) = 0.835$.

$VD<sequence>$

$= (VD<receive> + VD<switch> + VD<invoke_{s3}> + VD<reply>) / 4$

$= (0 + 0.835 + 0.33 + 0) / 4 = 0.29$.

The $<sequence>$ activity in $P$ is the root node and its derived variability degree denotes the whole process. Thus, $VD<sequence> = VD<root> = 0.29$. Hence we derive:

$\textbf{PAM}_P = VD<sequence_{root}> = \textbf{0.29}$.

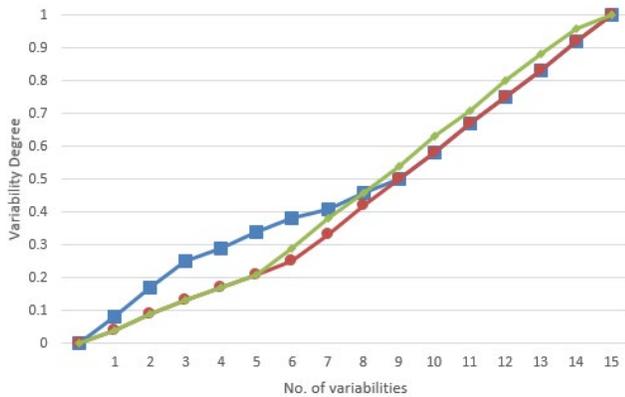

**Figure 6. Variability degrees of the travel booking process.**

In Figure 6, we show three cases of derived variability degrees of the travel booking process $P$. A line in the graph represents a case composed of a set of variability degrees that can be derived from $P$, considering at each point the number of variabilities. We incrementally specify the variabilities to any variable activities of $P$. For each case, we start by deriving a $PAM_P$ where $P$ has a single variability; then we repeat the derivation for each added variability until all variabilities are considered in $P$. With the five variable activities of $P$, we can specify a maximum of 15 variabilities, which will make the process achieve its full adaptability (i.e. $PAM_P = 1$). We observe in all three cases that the variability degrees of the process gradually increase as more variabilities are added. If we compare the variability degrees at every variability number, this is regardless of which join point (i.e. variable activity) a variability is specified in $P$, the derived set of values are relatively similar. However, we notice some slight differences in the variability degrees at a certain number of variabilities, such as when the number of variabilities in $P$ is six or seven. The difference in values is the result of the distinct placement of variabilities in the process, especially for a process with multiple hierarchical structured constructs.

## 6. CONCLUSION AND FUTURE WORK

In this paper, we propose metrics to quantify the adaptability of an AO4BPEL process. We examine the adaptability dimension that is addressed in the integration of the AOP paradigm with BPEL. Such adaptability changes the structural behaviour of a process to handle crosscutting concerns such as changes in business rules. The metrics can be an aid in the quality evaluation of a process. In addition, it can be helpful in the comparison of process configurations to achieve a desired adaptability degree. Although we focus on the AO4BPEL process, this work opens the opportunity to evaluate the adaptability of process specifications from the other aspect-oriented BPEL frameworks as well as with the non-aspect-oriented BPEL extensions.

Our further work will include validation of our approach in its applicability to processes specified from other aspect-oriented BPEL frameworks presented in the literature (e.g. BPELn'Aspects [4] and MODAR [18]). An implementation tool could also be developed to automate the computation. Likewise, we look into enhancing the metrics to include other adaptability dimensions such as binding variability which considers the adaptability in selecting partner services.

# 7. REFERENCES

[1] Alferez, G. H., Pelechano, V., Mazo, R., Salinesi, C., and Diaz, D. 2014. Dynamic adaptation of service compositions with variability models. *Journal of Systems and Software*, 91 (May 2014), 24-47. DOI=http://dx.doi.org/10.1016/j.jss.2013.06.034.

[2] Charfi, A., and Mezini, M. 2007. AO4BPEL: An aspect-oriented extension to BPEL. *World Wide Web*, 10, 3 (September 2007), 309-344. DOI=10.1007/s11280-006-0016-3.

[3] Jordan, D., Evdemon, J., Alves, A., Arkin, A., Askary, S., Barreto, C., Bloch, B., Curbera, F., Ford, M., and Goland, Y. 2007. Web Services Business Process Execution Language version 2.0. https://docs.oasis-open.org/wsbpel/2.0/PR02/wsbpel-specification-draft-diff.pdf.

[4] Karastoyanova, D., and Leymann, F. 2009. BPEL'n'Aspects: Adapting service orchestration logic. In *Proceedings of the ICWS 2009. IEEE International Conference on Web Services*. IEEE.

[5] Khoshkbarforoushha, A., Jamshidi, P., Gholami, M. F., Wang, L., and Ranjan, R. 2016. Metrics for BPEL process reusability analysis in a workflow system. *IEEE Systems Journal*, 10, 1 (March 2016), 36-45. DOI=10.1109/JSYST.2014.2317310.

[6] Khoshkbarforoushha, A., Jamshidi, P., Nikravesh, A., Khoshnevis, S., and Shams, F. 2009. A metric for measuring BPEL process context-independency. In *Proceedings of the 2009 IEEE International Conference on Service-Oriented Computing and Applications (SOCA)*. IEEE.

[7] Kiczales, G., Lamping, J., Mendhekar, A., Maeda, C., Lopes, C., Loingtier, J.-M., and Irwin, J. 1997. Aspect-oriented programming. In *Proceedings of the 11th European Conference on Object-Oriented Programming* (Finland). Springer.

[8] Lenhard, J. 2014. Towards quantifying the adaptability of executable BPMN processes. In *Proceedings of the 6th Central European Workshop on Services and their Composition* (Potsdam, Germany). Citeseer.

[9] Lenhard, J., Geiger, M., and Wirtz, G. 2015. On the measurement of design-time adaptability for process-based systems. In *Proceedings of the Symposium on Service-Oriented System Engineering (SOSE)*. IEEE.

[10] Metzger, A., Pohl, K., Papazoglou, M., Di Nitto, E., Marconi, A., Karastoyanova, D., Agarwal, S., Berre, A., Brogi, A., and Bucchiarone, A. 2012. Research challenges on adaptive software and services in the future internet: Towards an S-cube research roadmap. In *Proceedings of the Proceedings of the First International Workshop on European Software Services and Systems Research: Results and Challenges* (Zurich, Switzerland). IEEE Press.

[11] Mirandola, R., Perez-Palacin, D., Scandurra, P., Brignoli, M., and Zonca, A. 2015. Business process adaptability metrics for QoS-based service compositions. In *Proceedings of the European Conference on Service-Oriented and Cloud Computing*. Springer.

[12] Murguzur, A., Intxausti, K., Urbieta, A., Trujillo, S., and Sagardui, G. 2014. Process flexibility in service orchestration: A systematic literature review. *International Journal of Cooperative Information Systems*, 23, 3 (September 2014), 1430001. DOI=http://dx.doi.org/10.1142/S0218843014300010.

[13] Nguyen, T., Colman, A., and Han, J. 2011. Modeling and managing variability in process-based service compositions. *Service-Oriented Computing*, Springer, 7084, 404-420.

[14] Perez-Palacin, D., Mirandola, R., and Merseguer, J. 2014. On the relationships between QoS and software adaptability at the architectural level. *Journal of Systems and Software*, 87 (January 2014), 1-17. DOI=http://dx.doi.org/10.1016/j.jss.2013.07.053.

[15] Sheng, Q. Z., Qiao, X., Vasilakos, A. V., Szabo, C., Bourne, S., and Xu, X. 2014. Web services composition: A decade's overview. *Information Sciences*, 280 (October 2014), 218-238. DOI=http://dx.doi.org/10.1016/j.ins.2014.04.054.

[16] Subramanian, N., and Chung, L. 2001. Metrics for software adaptability. In *Proceedings of the Software Quality Management (SQM April 2001)* (Loughborough, UK).

[17] von Rosing, M., White, S., Cummins, F., and de Man, H. 2015. Business Process Model and Notation. http://www.omg.org/news/whitepapers/Business_Process_Model_and_Notation.pdf.

[18] Yu, J., Sheng, Q. Z., Swee, J. K. Y., Han, J., Liu, C., and Noor, T. H. 2015. Model-driven development of adaptive web service processes with aspects and rules. *Journal of Computer and System Sciences*, 81, 3 (May 2015), 533-552. DOI=10.1016/j.jcss.2014.11.008.